\documentclass[12pt]{iopart}
\input epsf
\newcommand{\be}{\begin{equation}}
\newcommand{\ee}{\end{equation}}
\newcommand{\bea}{\begin{eqnarray}}
\newcommand{\eea}{\end{eqnarray}}
\begin{document}
\title[Dynamics of proving uncolourability of large random graphs]
{The dynamics of proving uncolourability of large random graphs\\
I. Symmetric Colouring Heuristic}
\author{Liat Ein-Dor$\dag$  and R\'emi Monasson${\dag,\ddag}$}
\address{
\dag CNRS-Laboratoire de Physique Th\'{e}orique de l'ENS, 24 rue
Lhomond, 75005 Paris, France} \address{ \ddag  CNRS-Laboratoire de
Physique Th\'{e}orique, 3 rue de l'Universit\'{e}, 67000
Strasbourg, France}

\begin{abstract}
We study the dynamics of a backtracking procedure capable of
proving uncolourability of graphs, and calculate its average
running time $T$ for sparse random graphs, as a function of the
average degree $c$ and the number of vertices $N$. The analysis is
carried out by mapping the history of the search process onto an
out-of-equilibrium (multi-dimensional) surface growth problem. The
growth exponent of the average running time, $\omega(c)= (\ln T)/ N$, is
quantitatively predicted, in agreement with simulations.
\end{abstract}


\section{Introduction}

The wide variety of practical problems that can be mapped onto
NP-complete problems, together with the challenge in finding an
answer to one of the most important open questions in theoretical computer
science, `Does $NP=P?$', have led to intensive studies in the
past decades. Despite intense efforts, the worst case running times of
all currently known algorithms grow exponentially with the size of the
inputs to these problems. However, NP-complete problems are
not always hard. They might be even easy to solve on average
\cite{Garey,Cheeseman,Turner} {\em i.e.} when their
resolution complexity is measured with respect to some underlying
probability distribution of instances.  This `average-case'
behavior depends, of course, on the input-distribution.

In the graph colouring problem, one of the most well-known
combinatorial optimization problems with applications ranging from
time tabling and scheduling \cite{Leighton,Werra}, through
register allocation \cite{Briggs,Chow}, to frequency assignment
\cite{Meurdesoif}, the average-case behavior is often defined on
random graphs.  The aim is to colour the vertices of the graph such
that no adjacent vertices have the same colour.  Whether this can
be done with $k$ or less than $k$ colours constitutes the so called
$k$-colouring ($k$-COL) decision problem. $2$-COL is easy and can
be decided in a time growing polynomially with the size (number of
vertices) of the graph, while $k$-COL is NP-complete for any $k\ge
3$ \cite{Miller,Garey}. We shall restrict to the investigation of
3-COL in the following, and denote by Red (R), Green (G) and Blue
(B) the available colours. Graphs will be generated according to
the $G(N,p)$ distribution: they are made of $N$ vertices, linked
two by two through edges with probability $p$.  Studies of the
3-COL problem on this ensemble have indicated that, for sparse
random graphs in which $p=c/N$, a phase transition between
colourable and uncolourable phase occurs in the large $N$ limit
as the connectivity (average vertex--degree) $c$ is varied
$c$ \cite{Friedgut,Gent}. Below a critical value $c_3$ of the
connectivity $c$, almost all
instances are colourable whereas, above $c_3$, the probability that
an instance is colourable drops to zero. Determination of $c_3$ is
an open question in random graph theory first posed by
Erd\"os \cite{Alon}.  Nevertheless, years of investigation have
yielded some lower and upper bounds for $c_3$. Probabilistic
counting arguments have led to the best known upper bound
$c_3<4.99$ \cite{Kaporis}. A recent analysis of a ``smoothed''
version of the Brelaz heuristic \cite{Brelaz} has yielded the
highest lower bound $c_3>4.03$ \cite{Moore}.

In a recent work, Mulet {\em et al.} used a mapping of the graph
colouring problem onto the Potts model, and applied statistical mechanics
methods to estimate $c_3$ \cite{Mulet}.  The result, $c_3\approx
4.69$, is very close to numerical simulations
\cite{Boettcher}. Below $c_3$, solving 3-COL can be done by exhibiting
a proper colouring, a task carried out by search algorithms in an
apparently
efficient way \cite{Mulet}.  Above $c_3$, resolution of an instance
almost surely means exhibiting a proof of its uncolourability, a
very hard task. One of the most popular algorithm capable
of exhibiting such proofs is the Davis-Putnam-Logemann-Loveland procedure
(DPLL) \cite{Davis}. Its operation amounts to a clever exhaustive search in
the configuration space, based on the errors and trials principle.
Generally, the time needed by DPLL to check the absence of colouring grows
exponentially with the size of the graph, $T \sim \exp (N\, \omega (c) )$.
The purpose of this paper is to calculate $\omega$ as a function of $c$.
Such a study was recently undertaken for the satisfiability problem
\cite{Cocco,ptac} and vertex covering \cite{Weigt,ptac}, both hard decision
problems. The interest of 3-COL with respect to the latter cases is its
intrinsic symmetry. From any proper colouring, five other colourings
can be deduced through colour permutations. It is therefore interesting to
understand whether respecting or breaking this symmetry can lead to
computational gains, and how this can be implemented in the dynamics
of the search algorithm \cite{Asymmetric}.

Hereafter, we focus on the case of colouring heuristics that do
not explicitly break the symmetry between colours. The analysis of
the biased case is left to a forthcoming companion paper
\cite{Asymmetric}. This article is organized as follows. The
colouring algorithm is presented in section \ref{SecAlgorithm}.
Section \ref{SecBack} is devoted to the analysis of the dynamics
and of the resolution time of the algorithm. In the last section
\ref{SecSummary} we summarize and  propose some perspectives.

\section{Description of the Colouring Algorithm}
\label{SecAlgorithm}

The algorithm which we analyze in this paper is a complete algorithm
capable of determining whether a given graph is $3$-colourable or not.
The algorithm is
based on a combination of a colouring heuristic, $3$-GREEDY-LIST ($3$-GL),
and of backtracking steps. Its operation is exposed below.

\subsection{Operation of the Greedy-List algorithm with backtracking}

\begin{figure}
\begin{center}
\mbox{\epsfxsize 5.2in \epsfbox{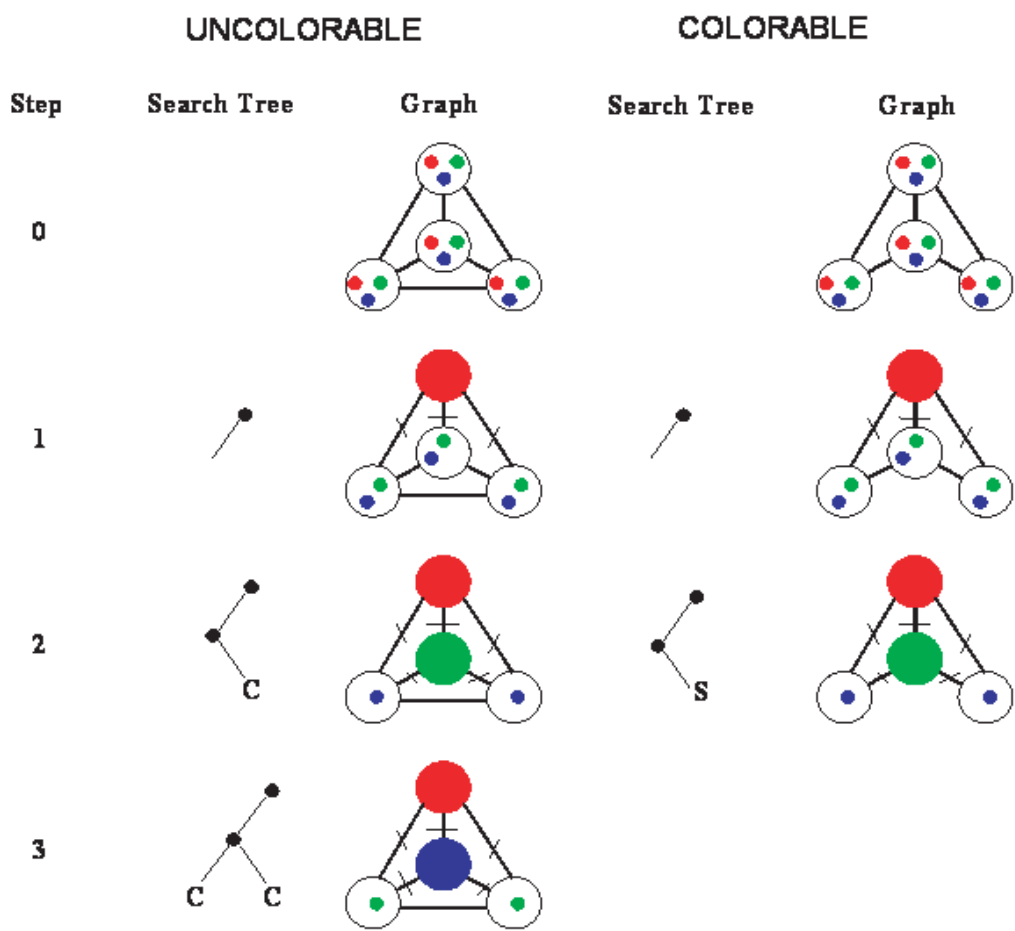}}
\vspace{1cm}
\end{center}
\caption{Two examples which demonstrate how the GL
algorithm acts onto a colourable (left side) and an uncolourable
(right side) graph. The figure illustrates how the search tree grows
with the operation of the algorithm. Available colours at each step
are denoted by the patterns of the filled circles attached to
vertices. When a vertex is coloured, it is removed from the graph,
together with all its attached edges. In addition, the chosen colour
is removed from the neighours' sets of available colours. On the left
side of the figure, a colourable graph is coloured by the
algorithm. No contradiction is encountered, and the algorithm finds a
solution without backtracking. On the right side, the algorithm tries
to colour an uncolourable graph. When it first hits a contradiction
(step 2) {\em i.e.} when two 1-colour vertices connected by an edge
are left with the same available colour, the algorithm backtracks to
the last-coloured vertex, and tries to colour it with the second
available colour. When a contradiction is hit again, the algorithm
terminates. Note, that in principle, it could backtrack to the
first-coloured node, and try other colour options. However, due to
colour gauge symmetry, this will not yield a solution.}
\label{example}
\end{figure}

The action of the colouring procedure is illustrated on Figure
\ref{example} and described as follows:

\begin{itemize}
\item Necessary Information: while running, the algorithm maintains
      for each uncoloured vertices, a list of available colours, which
      consists of all the colours that can be assigned to this vertex
given the colours already assigned to surrounding vertices.

\item Colouring Order: the order in which the vertices are coloured,
      is such that the most constrained vertices {\em i.e.} with the least
      number of available colours are coloured first.
      At each time step, a vertex is chosen among the most
      constrained vertices, and its
      colour is selected from the list of its available colours.
      Both choices are done according to some heuristic rule, which can
      be unbiased (no preference is made between colours), or
      biased (following a hierarchy between colours), see next Section.

\item List-Updating: to ensure that no adjacent vertices have the same colour,
      whenever a vertex is assigned a colour, this colour is
      removed from the lists (if present)
      attached to each of the uncoloured neighbors.

\item Contradictions and Backtracking: a contradiction occurs as soon
      as one of the lists becomes empty. Then, the algorithm backtracks to
      the most recently chosen vertex, which have more than one available
      colour (the closest node in the search tree - see
      definition below).

\item Termination Condition: the algorithm stops when all vertices
      are coloured, or when all colouring possibilities have been tried.
\end{itemize}

A search tree describes the action of the algorithm is the following, with
the following components:

\begin{itemize}
\item  Node: a node in the tree represents a vertex chosen by the algorithm,
       which has more than one colour in its available-colours-list.

\item  Edge: an edge which comes out of a node, corresponds to a
       possible colour of the chosen vertex.

\item  Leaf: a branch terminates either by a solution (denoted by $S$) or by
       a contradiction (denoted by $C$), depending on whether the colour
choices
       made along this branch give a proper colouring of the graph, or not.
\end{itemize}

\subsection{Colour symmetry: the unbiased 3-GL heuristic}
\label{important}

Let us call 3-GL heuristic the incomplete version of the above algorithm,
obtained when the algorithm stops
if a colouring is found (and outputs ``Colourable''), or just after the
first contradiction instead of backtracking (and outputs ``Don't know
if colourable or not''). In contrast to 3-GL algorithm with backtracking,
the 3-GL heuristic is not able to prove the absence of solution, and is
amenable to rigorous analysis \cite{Moore}.

In the simplest case, vertices and colours are chosen purely randomly
without any bias between colours (Colouring Order step described above).
This `symmetric' 3-GL heuristic verifies two key properties
which our analysis rely on. The first one is a statistical invariance
called R-property.
Throughout the execution of the algorithm, the uncoloured part of the
graph is distributed as $G((1-t)N,p)$ where $t$ is the number of
coloured vertices divided by $N$.  The second property is
colour symmetry.  The search heuristic is symmetric with respect to the
different colours, and the initial conditions are symmetric as well.
Denoting by $l=\{R,G,B\}$ the list of the three available colours, a
$2$-colour node can have one of three possible lists
$\{R,G\}$,$\{R,B\}$,$\{G,B\}$ and similarly, there are three
possible lists for a $1$-colour node.  Due to colour symmetry, in
the limit of large $N$, we expect the groups of $1$-colour and $2$-colour
vertices to be composed of an equal number of vertices (with $o(N)$
fluctuations) with the three kinds of lists. Hence, in the
leading order, the evolution of the algorithm can be expressed by the
evolution of the three numbers $N_{j}(T)$ of $j$-colour nodes ($j=1,2,3$).
The analysis of the evolution of these numbers in the course of the
colouring was done by Achlioptas and Molloy\cite{Ach}. It is briefly
recalled below.

\subsection{Analysis of the symmetric 3-GL heuristic}
\label{SecNoBack}

In the absence of backtracking, 3-GL terminates as soon
as a contradiction occurs, or a solution is found.
Differential equations can be used to track the evolution of node populations
as colouring proceeds \cite{Moore}. In this section we briefly recall
how to obtain these differential equations, and the associated search
trajectories of the heuristic in terms of node populations.

According to the R-property, the probability that an $j$-colour node
is a neighbour of the currently coloured node equals $c/N$ throughout the
running of the heuristic. The
probability that the same colour appears in its list is $j/3$.
Therefore the two average flows of vertices, $w_2(T)$ from $N_{3}(T)$ to
$N_{2}(T)$, and $w_1(T)$ from $N_{2}(T)$ to $N_{1}(T)$ are $c\,N_3(T)/N$ and
$2\,c\,N_2(T)/(3\,N)$ respectively.
Hence, the evolution equations for the three populations of vertices
read,
\bea
N_3(T+1) &=& N_3(T)-w_2(T) \quad , \nonumber \\
N_2(T+1) &=& N_2(T)+w_2(T)-w_1(T)-\delta N_1(T) \quad , \nonumber \\
N_1(T+1) &=& N_1(T)+w_1(T)-(1-\delta N_1(T)) \quad . \eea where
$\delta N_1(T)=1$ if $N_1(T)=0$ (a $2$-colour vertex is coloured)
and $\delta N_1(T)=0$ if $N_1(T)\ne 0$  (a $1$-colour vertex is
coloured). For $c>1$, both $N_2(T)$ and $N_3(T)$ are extensive in
$N$, and can be written as \be N_i(T)=n_i(T/N)\;N+o(N)\quad .
\label{scaling} \ee Apparition of the reduced time, $t=T/N$, means
that population densities $n_i(T/N)$ change by $O(1)$ over $O(N)$
time intervals. To avoid the appearance of contradictions, the
number of $1$-colour vertices must remain of $O(1)$ throughout the
execution of the algorithm. From queueing theory, this requires
$w_1(t)<1$, that is \be \frac 23\; c\; n_2(t) < 1
\label{condition} \ee which means that $1$-colour nodes are
created slowly enough to colour them and do not accumulate. Thus,
in the absence of backtracking, the evolution equations for the
densities are \be \frac{dn_3(t)}{dt} = -c\;n_3(t) \ , \quad
\frac{dn_2(t)}{dt} = c\; n_3(t)-1\ . \label{dn23} \ee The solution
of these differential equations, with initial conditions
$n_3(0)=1$, $n_2(0)=0$, is \be n_3(t) = e^{-c\,t} \ ,\quad n_2(t)
= 1-t - e^{-c\,t}\ . \label{n23} \ee Eqs. (\ref{dn23}) were
obtained under the assumption that $n_2(t)>0$ and hold until
$t=t_2$ at which the density $n_2$ of 2-colour nodes vanishes. For
$t>t_2$, $2$-colour vertices do not accumulate anymore. They are
coloured as soon as they are created. $1$-colour vertices are
almost never created, and the vertices coloured by the algorithm
are either $2$-, or $3$-colour vertices. Thus, when $t_2<t<1$,
$n_2(t)=0$, and $n_3(t)=1-t$ decreases to zero. A proper coloring
is found at $t=1$ {\em i.e.} when all nodes have been coloured.

\begin{figure}[ht]
\begin{center}
\mbox{\epsfxsize 3.5in \epsfbox{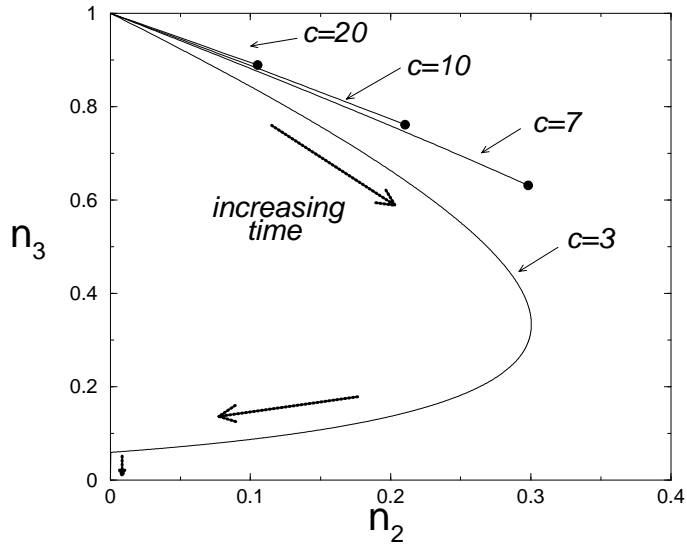}}
\end{center}
\caption{Trajectories of dominant search branches generated by the
$3$-GL in the UNCOL phase ($c>c_3\simeq 4.7$). compared to a
search trajectory in the easy COL phase ($c<c_L\simeq 3.85$).
Horizontal and vertical axis represent the densities $n_2$ and
$n_3$ of $2$- and $3$-colour nodes respectively. Trajectories are
depicted by solid curves, and the arrows indicate the direction of
motion (increasing depth of the search tree); they originate from
the left top corner, with coordinates $(n_2=0,n_3=1)$, since all
nodes in the initial graph are $3$-colour nodes.  Dots at the end
of the UNCOL trajectories ($c=7,10,20$) symbolize the halt point
at which condition $n_2 < 3 \ln 2/c$ ceases to be fulfilled, and
the search tree stops growing (\ref{posi}). Note that as the
initial connectivity increases, the trajectories halt at earlier
stage, implying the early appearance of contradictions as the
problem becomes over-constrained (large connectivity values).  The
COL trajectory (shown here for $c=3$) represents the
under-constrained region of the problem, where the very first
search branch is able to find a proper colouring is found (bottom left
corner with coordinates ($n_2=0,n_3=0$)).
}\label{diagram}
\end{figure}

These equations define the trajectory of the algorithm in phase
space in the absence of contradictions {\em i.e.} as long as condition
(\ref{condition}) is fulfilled. The trajectory corresponding to $c=3$ is
plotted on Figure~\ref{diagram}. For $c<c_L\approx 3.847$,
condition (\ref{condition}) is never violated, and the probability
that the algorithm succeeds in finding an appropriate
colouring without backtracking is positive.
The complexity $\gamma (c)\, N$ of the
algorithm in this regime of $c$ is linear with $N$, and equals
the number of nodes in the single branch of the search tree.
\be
\gamma (c)= 1-\frac 23 \; c\int_0^{t^*} dt\;n_2(t) \ ,
\ee
where $t^*>0$ is the first time (after $t=0$) that $n_2(t)$ becomes $0$.

For $c>c_L$ condition (\ref{condition}) is violated at $t=t_d(c)$ which
depends on $c$, and $1$-colour vertices start to accumulate. As a
result, the probability for contradictions becomes large,  and
backtracking enters into play.

\section{Study of the 3-Greedy List algorithm with backtracking}
\label{SecBack}
The analytical study of the complexity in the presence of
backtracking is inspired from previous analysis of random 3-SAT solving
with DPLL algorithm \cite{Cocco,Coccotcs}.

\subsection{Evolution equation for the search process}

\begin{figure}[ht]
\begin{center}
\mbox{\epsfxsize 4in \epsfbox{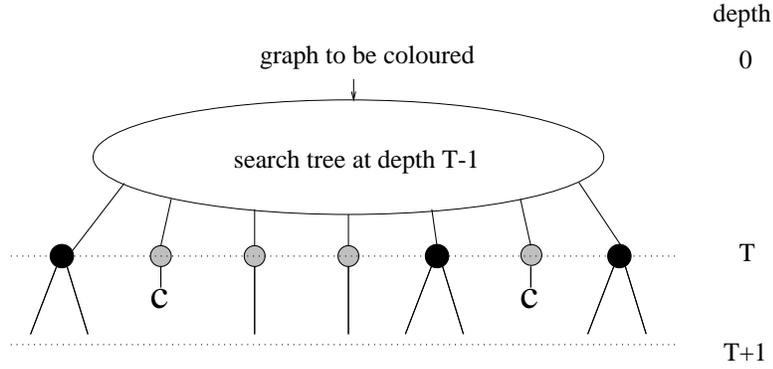}}
\end{center}
\caption{Imaginary, breadth--first growth process of a search tree associated
to an UNCOL graph and used in the
theoretical analysis. $T$ denotes the depth in the
tree, that is the number of nodes coloured along each
branch. At depth $T$, one node is chosen on each branch among 1-colour
vertices if any (grey circles), or 2-, 3-colour
(splitting, black circles).
If a contradiction occurs as a result of 1-colour node colouring,
the branch gets  marked with C and dies out. The growth of the tree proceeds
until all branches carry C leaves. The resulting tree is identical to the one
built through the usual, sequential operation of the 3-GL algorithm.}
\label{struct}
\end{figure}

In the absence of solution, the algorithm builds a complete search tree
before stopping. In a complete tree $Q+1=B$, where $B$ is the number of
leaves and $Q$ the number of nodes. This relation implies that
the key for obtaining the complexity $Q$ lies in the calculation of $B$.
In order to enable a mathematical analysis of $B$, we rely on the fact
that the search tree is complete, and therefore the sequential
(depth-first) way in which the algorithm
builds it is irrelevant to the final structure. In other words, the order in
which the available colours of a vertex are tried, does not affect
the final shape of the tree. An identical tree can be built in a
parallel (breadth-first) way defined as follows, and illustrated in
Figure~\ref{struct}.

At time $T=0$, the tree reduces to a root node, to which is
attached the graph to colour, and an attached outgoing edge. At
time $T$, that is, after having coloured $T$ vertices of the
graphs attached to each branch, the tree is made of $\tilde B(T) \
(\le 2^T)$ branches, each one carrying a partially coloured graph.
At next time step $T\to T+1$, a new layer is added to the tree by
colouring, according to 3-GL heuristic, one more vertex along
every branch. As a result, at each instant of the parallel
process, branches either die (encounter a contradiction), keep
growing (a $1$-colour vertex is coloured), or split (a $2$-colour
vertex is chosen and its two available colours are tried
simultaneously) (Figure~\ref{struct}).  This parallel growth
process is Markovian, and can be encoded in an instance--dependent
(and exponentially large in $N$) evolution matrix $H$
\cite{Coccotcs}.

To obtain a tractable expression for $H$, we neglect correlations arising from
the choice of the same vertex in two different branches.
After assigning $T$ variables, each branch
represents a different sequence of $T$ coloured vertices, which
determines the values $\{N_1,N_2,N_3\}$ attached to this branch.
Denoting by $\tilde B(N_1,N_2,N_3;T)$ the
number of branches at time $T$ with $N_i$ ($i=1,2,3$) $i$-colour
vertices, the growth process of the search tree can be described
by the evolution of $\tilde B(N_1,N_2,N_3;T)$ with time. This
evolution is given by
\be \fl \tilde
B(N_1,N_2,N_3;T+1)=\sum_{N_1',N_2',N_3'=0}^\infty
H(N_1,N_2,N_3,N_1',N_2',N_3';T)\;\tilde B(N_1',N_2',N_3';T),
\label{evolution} \ee where \bea \fl
H(N_1,N_2,N_3,N_1',N_2',N_3';T)=\sum_{w_2=0}^{N_3'}
{{N_3'}\choose{w_2}} ({{c}\over
{N}})^{w_2}(1-{{c}\over{N}})^{N_3}
\delta_{N_3'-N_3-w_2} \times \nonumber \\
\fl \left\{(1-\delta_{N_1'})\sum_{w_1=0}^{N_2'}{{N_2'}\choose{w_1}}
({{2c}\over{3N}})^{w_1}
(1-{{2c}\over{3N}})^{N_2'-w_1}\delta_{N_2-N_2'-(w_2-w_1)}
\delta_{N_1-N_1'-w_1+1}+\nonumber \right.\\
\fl\left. 2\delta_{N_1'}
\sum_{w_1=0}^{N_2'-1}{{N_2'-1}\choose{w_1}}({{2c}
\over{3N}})^{w_1}(1-{{2c}\over{3N}})^{N_2'-w_1-1}
\delta_{N_2-N_2'-(w_2-w_1-1)}\delta_{N_1-N_1'-w_1}\right\}
\label{K} \eea is the branching matrix of the $3$-GL algorithm,
and $\delta N$ is the Kronecker delta function. The matrix
describes the average number of branches with $\{N_i\}_{i=1}^3$
$i$-colour vertices, which are coming out from branches with
$\{N_i'\}_{i=1}^3$ $i$-colour vertices, as a result of all the
colouring options of the vertex coloured at time $T$. The
R-property is responsible for the binomial distributions of the
flows $w_1$ and $w_2$ in (\ref{K}). Note that (\ref{K}) is written
under the assumption that no $3$-colour nodes are chosen by the
algorithm throughout the growth process. This assumption is
consistent with the resultant solution which shows that in the
uncolourable (UNCOL) region, $n_2(t)$, namely the number of
2-colour vertices divided by $N$, keeps positive for all $t>0$.

\subsection{Resolution of the evolution equation}
\label{Sres}

In order to obtain the complexity from the evolution equation of
$\tilde B(\vec{N};T)$ (\ref{evolution}), we define the generating
function $B(\vec{y};T)$ of $\tilde B(\vec{N};T)$ to be
\be \fl B(\vec{y};T)=\sum_{\vec{N}}\exp(\vec{y}\bullet
\vec{N})\tilde B(\vec{N};T), \; \; \; \vec{y}=(y_1,y_2,y_3), \;
\vec{N}=(N_1,N_2,N_3). \label{generating}
\ee
Plugging (\ref{K},\ref{generating}) into
(\ref{evolution}) yields the following evolution equation for the
generating function $B(\vec{y};T)$,
\be \fl
B(\vec{y};T+1)=e^{-y_1}\;B\big(\vec g(\vec{y});T\big)+\big(2\;
e^{-y_2}-e^{-y_1}\big) \;B\big(-\infty,g_2(\vec{y}),g_3(\vec{y});T\big),
\label{gen_evolution} \ee where \bea
g_1(\vec{y}) &=& y_1 \quad ,\nonumber \\
g_2(\vec{y}) &=& y_2+{{2c}\over{3N}}\big( e^{y_1-y_2}-1\big) \quad
, \nonumber \\ g_3(\vec{y}) &=& y_3+{{c}\over{N}}\big(
e^{y_2-y_3}-1\big) \ . \eea To solve (\ref{gen_evolution}), we
make scaling hypothesis for $\tilde B$ and $B$ in the large $N$
limit\cite{Cocco}. Let us examine how a step of the algorithm
affects the size of the three populations $N_1,N_2,N_3$. Since the
average connectivity is $O(1)$ {\em i.e.} each vertex is connected
on average only to $O(1)$ vertices, when a vertex is coloured, the
number of vertices whose status (the number of available colours)
is subsequently changed is bounded by the number of neighbors of
the coloured vertex. Hence a reasonable assumption is that the
densities $n_i=N_i/N$ change by $O(1)$ after $T=t\times N$
vertices are coloured. The corresponding Ansatz for the number of
branches is, \be  \tilde B(\vec{N};T)=e^{N\, \omega(n_1,n_2,n_3;t)
+ o(N)} \label{hypo} \ee where non-exponential terms in $N$ depend
on the populations of $i$-colour nodes ($i=1,2,3$). From
(\ref{hypo}) and (\ref{generating}) we obtain the following
scaling hypothesis for the generating function $B$, \be
B(\vec{y};T)=e^{N\,\varphi(y_1,y_2,y_3;t)+o(N)}, \label{Anatz} \ee
where $\varphi(\vec{y};t)$ is the Legendre transform of
$\omega(\vec{n};t)$, the logarithm of the number of branches
$B(\vec{N};T)$ divided by $N$, \bea \varphi(\vec y;t)&=&
\max_{\vec n}\left[\omega(\vec n;t)+\vec y\cdot \vec n\right]
\nonumber \\\omega(\vec n;t)&=& \min_{\vec y}\left[\varphi(\vec
y;t)-\vec y\cdot \vec n \right] \label{Leg}
 \eea
where $\vec n=(n_1,n_2,n_3)$. $\vec y$ and $\vec n$ are conjugated
Legendre variables; in particular, the typical fraction of $i$-colour nodes
at depth $t$ are given by the derivatives of $\varphi$ at vanishing argument,
\be
n_i (t) = \frac{\partial \varphi}{\partial y _i} (\vec y =0;t)
\quad . \label{der}
\ee
At the initial stage of the tree building up, there is a single outgoing
branch from the root node, carrying a fully uncoloured graph. Thus,
$\tilde B(\vec N;T=0)=1$ if $\vec N = (0,0,N)$, $0$ otherwise,
and $B(\vec y, T=0)= e^{N\, y_3}$. The initial condition for function
$\varphi$ is simply,
\be
\varphi (\vec y; t=0) = y_3 \quad .
\label{init}
\ee
According to (\ref{scaling}) both $N_2(T)$ and $N_3(T)$ are
extensive in $N$; hence $n_2>0$ and $n_3>0$.
Conversely, as soon as $N_1(T)$ becomes very large,
contradictions are very likely to occur, and the growth process stops.
Throughout the growth process, $N_1=O(1)$ almost surely.
Thus $n_1=0$ with high probability, and $\varphi$ does not depend upon
$y_1$ from (\ref{der}).

Independence of $\varphi$ on $y_1$ allows us to choose the latter
at our convenience, that is, as a function of $y_2,y_3,t$.
Following the so-called kernel method \cite{knu}, we see that
equation (\ref{gen_evolution}) simplifies if $y_1=y_2-\ln 2$.
Then, from Ansatz (\ref{Anatz}), we obtain the
following partial differential equation (PDE),
\be
\fl \frac{\partial\varphi}{\partial t}(y_2,y_3;t)=-y_2+\ln 2 -{{c}\over 3}
\frac{\partial\varphi}{\partial y_2}(y_2,y_3;t)+c\;(e^{y_2-y_3}-1)
\frac{\partial\varphi}{\partial y_3}(y_2,y_3;t)\ .
\label{partial}
\ee
The solution of PDE (\ref{partial}) with initial condition (\ref{init})
reads
\be \fl
\varphi(y_2,y_3;t)=
\frac{c}{6}t^2-\frac{c}{3}t+(1-t)(y_2-\ln2)+
\ln \bigg[3 +\;e^{-2 c\, t/3} \bigg(2\;e^{y_3-y_2} -3\bigg) \bigg] \ .
\label{phi} \ee

\subsection{Growth process of the search tree}
\label{SecPS}

\begin{figure}
\begin{center}
\mbox{\epsfxsize 4in \epsfbox{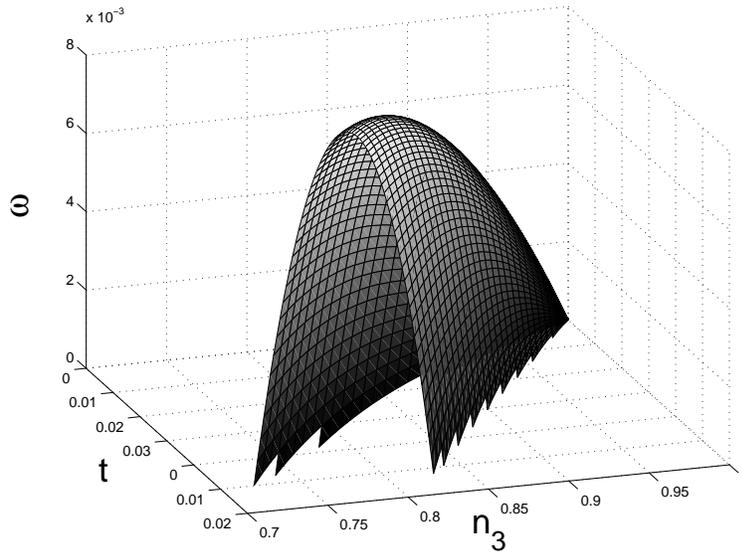}}
\end{center}
\caption{Function $\omega$ (log. of number of branches with
densities $n_2=1-t-n_3$, $n_3$ of 2- and 3-colour nodes at depth
$t$ in the search tree) as a function of $n_3$ and $t$ for $c=10$.
The top of the curve at given time $t$, $\omega ^*(t)$, is reached
for the dominant branch 3-colour density $n_3^*(t)$.
  The evolution of $\omega$ is shown till $t=t_{h}$ at which dominant
  branches in the search tree
  stop growing (die from the onset of contradictions). The maximal
$\omega$ at $t_{h}$, $\omega^*(t_h)$,
  is our theoretical prediction for the complexity.}\label{n2_n3}
\end{figure}

PDE (\ref{Anatz}) can be interpreted as a description of the
growth process of the search tree resulting from the algorithm
operation. Using Legendre transform (\ref{Leg}), PDE (\ref{Anatz})
can be written as an evolution equation for the logarithm
$\omega(n_2,n_3;t)$ of the average number of branches with
densities $n_2,n_3$ of 2- ,3-colours nodes as the depth $t=T/N$
increases, \be \frac{\partial \omega } {\partial t}  =
\frac{\partial \omega } {\partial n_2}  +\ln 2 -{{c}\over 3}
\;n_2 + c\; n_3 \left[ \exp \left(
\frac{\partial\omega}{\partial n_3}-
\frac{\partial\omega}{\partial n_2}\right) -1 \right] \ .
\label{comega} \ee The surface $\omega$, growing with ``time'' $t$
above the plane $n_2,n_3$ describes the whole distribution of
branches. Here, this distribution simplifies due to nodes
conservation. The sum $n_2+n_3$ of 2- and 3-colour nodes densities
necessarily equals the fraction $1-t$ of not-yet coloured nodes.
Therefore, $\omega$ is a function of $n_3$ and $t$ only, whose
expression is obtained through inverse Legendre transform of
(\ref{phi}), \bea \omega(n_3;t)&=&\frac {c}6\; t\; (1-2\,t-4\,
n_3) -n_3 \,\ln\, n_3 -
(1-n_3)\, \ln \,(1-n_3) - \nonumber \\
&& (1-t-n_3) \ln 2+(1-n_3)\, \ln \bigg[ 3 \;\bigg(1- e^{-\, 2
\,t\, c/3} \bigg) \bigg] \ . \eea
Figure \ref{n2_n3} exhibits $\omega(n_3,t)$ for $c=10$.

The average number  of branches at depth $t$ in the tree equals
\begin{equation}
\tilde B(t) = \int  dn_2\; dn_3 \; e^{N\, \omega
(n_2,n_3;t)} \sim e^{N\, \omega ^* (t)} \quad ,
\end{equation}
where
\be
\omega ^* (t) =
\frac{c}{6}t^2-\frac{c}{3}t -(1-t)\ln2+
\ln \bigg[3 - \;e^{-2 c\, t/3}\bigg]
\ee
is the maximum over $n_2,n_3$ of $\omega (n_2,n_3;t)$
reached in $n_2^*(t), n_3^*(t)$.
In other words, the exponentially dominant contribution to $\tilde B(t)$
comes from branches carrying partially coloured graphs with densities
\be
n^*_3(t)=  \frac{2}{3\;e^{\,2\, c\, t/3}-1}
\ , \quad
n^*_2(t) = 1-t-n^*_3(t) \ .
\ee
Under the action of the
$3$-GL algorithm, initially random $3$-colouring instances become
random mixed $2\&3$-colouring instances, where nodes can have either 2
or 3 colours at their disposal.  This phenomenon indicates that the
action of the $3$-GL algorithm on random $3$-colouring instances can be
seen as an evolution in the $n_2,n_3$ phase-space (Figure \ref{diagram}).
Each point $(n_2,n_3)$ in this space, represents a random mixed
$2\&3$-colouring instance, with an average number $(n_2+n_3)N$ of
nodes, and a fraction $n_3/(n_2+n_3)$ of $3$-colour nodes.  Parametric
plot of $n_2^*(t),n_3 ^*(t)$ as a function of $t$ represents the
trajectories of dominant branches in Figure~\ref{diagram}.

The search tree keeps growing as long as no contradictions are
encountered {\em i.e.} as long as $1$-colour vertices do not
accumulate.  This amounts to say that dominant branches are not
suppressed by contradictions and become more and more numerous through
2-colour nodes colouring,
 \be \frac{d \omega ^*}{dt}(t) > 0 \quad
, \label{posi} \ee or equivalently from (\ref{comega}), $n_2^*(t)<
3\ln 2/c$.  This defines the halt condition for the dominant branch
trajectories in the $n_2,n_3$ dynamical phase diagram of
Figure~\ref{diagram}. Call $t_h$ the halt time at which condition
(\ref{posi}) gets violated.  The logarithm $\omega^*(t_h)$ of the
number of dominant branches at $t=t_h$, when divided by $\ln 2$,
yields our analytical estimate for the complexity of resolution, $\ln
Q/N$.\footnote{Let us stress that our calculation is
approximate. First, as mentioned above, correlations between different
branches have been neglected.  Secondly, $\varphi$ is the Legendre
transform of $\omega$ over non-negative values of $n_i$ only, a
constraint we have not taken into account in the growth PDE
(\ref{partial}). We expect our prediction to be accurate when $n_2$
and $n_3$ are not getting to small throughout the growth process {\em
i.e.} for large enough connectivites $c$.}

\subsection{Comparison with numerical experiments}

\begin{figure}[ht]
\begin{center}
\mbox{\epsfxsize 3in \epsfbox{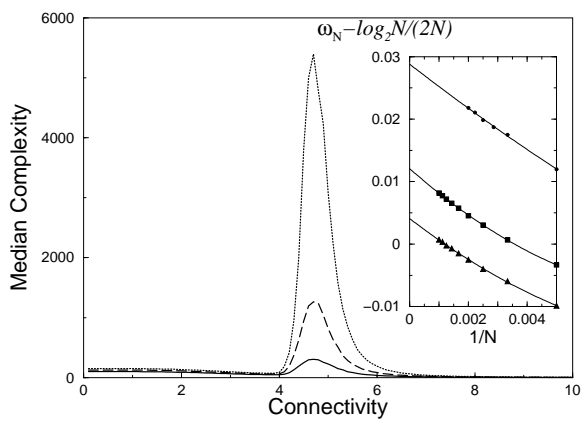}}
\end{center}
\caption{Simulation results exhibiting the easy-hard-easy pattern
which characterizes the complexity of the GL-algorithm. The curves
describe the median complexity as a function of connectivity, as
measured for $N=100$ (solid-line), $N=125$ (long-dashed line) and
$N=150$ (dotted-line), averaged over $1,000$ samples. The arrow
denotes the location of critical connectivity $c_3\sim4.7$
separating COL (left) from UNCOL (right) phases. Running times $T$
scale exponentially in the UNCOL phase, $T\sim 2^{\, N\,\omega}$.
Calculation of $\omega$ as a function of the connectivity $c$ in
the UNCOL phase is the purpose of the present work.
 Inset: polynomial fits (solid lines) to simulation
results of $\omega_{HIS}-\log_2N/(2N)$ vs. $1/N$ for
three different connectivity values $c=7$ (circles) $c=10$
(squares) and $c=15$ (triangles). The fits
are to $\omega_{HIS}-log_2N/(2N)$, to account for the
non-polynomial finite-size corrections to our saddle
point calculation. Extrapolations of the fits to the y-axis are our
estimates
for $\omega$ at $N\to \infty$, and appear in the second column of
Table~\ref{symtable}.
 Note that due
to high computational times, results for $c=7$ have been obtained
for sizes up to
$N=500$ only, and therefore provide a less accurate estimate of
$\omega$.}

 \label{sym_asym}
\end{figure}

To check our theory, we have run numerical experiments to estimate
$\omega$, the logarithm of the median solving time, as a function
of the initial graph degree $c$. Figure \ref{sym_asym} describes
the output of these simulations. The easy-hard-easy pattern of the
GC problem when passing from the COL ($c<c_3$) to the UNCOL
($c>c_3$) regions is clearly visible, with an exponential scaling
of hardness around the critical connectivity.

Table 1 presents results for $\omega$ as a function of the
connectivity $c$ in the UNCOL phase as found from numerical
experiments and from the above theory.  Note the significant
decrease in the complexity as the initial connectivity increases.
Extrapolation of numerical results to the large $N$limit is
described in the Inset of Figure \ref{sym_asym}. For $c=7$, the
agreement between numerical and analytical results is not perfect.
However, the high
computational complexity of the algorithm for small $c$ values, does
not allow us to obtain numerical results for large sizes $N$, and
affect the quality of the large $N$ extrapolation of $\omega$.

In the UNCOL region, as $c$ increases, contradictions emerge in an
earlier stage of the algorithm, the probability that the same vertex
appears in different branches reduces, and the analytical prediction
becomes exact. As a consequence of the early appearance of
contradictions, the complexity $\omega$ decreases with $c$.  At very
large $c$, we find
\be \omega (c)\asymp
\frac {3\, \ln 2}2 \; \frac 1{c^2} \simeq \frac{1.040}{c^2} \quad ,
\ee
and therefore that the (logarithm of the) complexity exhibits a
power law decay with exponent $2$ as a function of connectivity $c$.

\begin{table}
\begin{indented}
\item[]
  \begin{tabular}{|c|c|c|c|c|}
  \br
  $c$ & $\omega_{THE}$ & $\omega_{HIS}$ & $\omega_{NOD}$  \\
  \mr
   20 & $2.886*10^{-3}$ & $2.568*10^{-3}\pm5.85*10^{-4}$ & $3.038*10^{-3}\pm3.2*10^{-4}$ \\
   15 & $5.255*10^{-3}$ & $4.*10^{-3}\pm7.09*10^{-4}$ & $5.776*10^{-3}\pm4.79*10^{-4}$  \\
   10 & $1.311*10^{-2}$ & $1.371*10^{-2}\pm1.1*10^{-3}$ & $1.492*10^{-2}\pm9.6*10^{-4}$\\
   7  & $2.135*10^{-2}$ & $2.879*10^{-2}\pm6.8*10^{-3}$ & $3.091*10^{-2}\pm3.6*10^{-3}$\\
   \br
  \end{tabular}
\end{indented}
\caption{Analytical results and simulation results of the
complexity $\omega$ for different connectivities $c$ in the UNCOL
  phase.  The analytical values of $\omega_{THE}$ are derived from
  theory; $\omega_{HIS}$ is obtained by measuring the
  maximal number of branches in the histogram of branch
lengths\cite{Cocco}, and $\omega_{NOD}$ through direct measure of
the search tree size.\label{symtable}}
\end{table}

\section{Summary and Discussion}
\label{SecSummary}

In this study we have presented an analysis of the complexity of
the $3$-Greedy List (GL) algorithm acting onto uncolourable
(UNCOL) random-graph instances. This analysis provides an estimate
of the typical performances of the GL algorithm. Above the
colourability threshold $c_3$, proving the absence of colouring
takes a time growing exponentially with the size $N$ of the graph.
However, well above the threshold {\em i.e.} for graph
connectivities $c \gg c_3$, instances are strongly
over-constrained, and the absence of proper colouring is
established more and more quickly.  Complexities in this region,
though exponential with $N$, have a very small prefactor which for
large values of $c$ vanishes with a power law behaviour
($\omega(c)\asymp 1.040/c^2$).

The present study could be pursued in many directions.  First, it
would be interesting to analyse the performances of the 3-GL
algorithm in the colourable (COL) phase $c<c_3$.  Graphs with low
connectivities ($c<c_L$) being almost surely coloured in a time
growing linearly with their size \cite{Ach}, the interesting
region is the intermediate range of connectivities, $c_L<c<c_3$.
There, proper colourings are found at the cost of an exponential
computational effort, which could in principle be quantitatively
characterized along the lines of Ref.\cite{Cocco}.  Secondly,
another interesting extension would be to focus on other search
heuristics.  Attractive candidates are heuristics that favor
high-degree vertices. Such a procedure has been recently analyzed
(in the absence of backtracking) to improve rigorous lower bounds
to the COL-UNCOL threshold $c_3$ \cite{Moore}.  Last, the study of
more realistic e.g. finite dimensional graph distributions, could
aid in the understanding of the influence of instance structure on
resolution complexity.

As stated in the introduction, the main interest of 3-COL with
respect to other NP-complete problems e.g. SAT lies in its global
gauge symmetry. The 3-GL heuristic we concentrated on here does
not break this symmetry in that it treats on a equal foot all
2-colours nodes when a split has to made, irrespectively of their
attached list of available colors e.g. $\{R,G\}$, $\{R,B\}$ or
$\{G,B\}$. It is easy to design heuristics that explicitly
violates this symmetry and  preferentially colour nodes with $R$
if possible. The analysis of the computational performances of
backtracking algorithms based on such an asymmetric heuristic is
technically quite difficult, and will be presented in a
forthcoming work \cite{Asymmetric}.

A promising outcome of the present work is the relative technical
simplicity of our 3-GL analysis with respect to the corresponding
studies of DPLL on random SAT instances. The growth partial
differential equation monitoring the evolution of the search tree
could be solved exactly, in contradistinction with previous
studies of the SAT problem. This essentially comes from a simple
conservation law, the sum of the numbers of coloured and
uncoloured nodes remaining of course constant throughout the
search, and makes 3-GL with backtracking a good candidate for
future rigorous studies \cite{Coccotcs}.

\vskip 1cm {\bf Acknowledgements}

We would like to thank Simona Cocco for helpful discussions.
This work was supported in part by the ACI
Jeunes Chercheurs ``Algorithmes d'optimisation et syst\`emes quantiques
d\'esordonn\'es''. L.E. acknowledges financial support from
the post-doctoral funding program of the French Ministry of Research.

\end{document}